\documentclass{article}
\usepackage{dcase2021,amsmath,graphicx,url,times,booktabs, tabularx, enumitem, multirow,amssymb}
\usepackage[table,xcdraw]{xcolor}
\usepackage{array, makecell}
\def \am [#1]{\textcolor{red}{AM: #1}}

\title{Low-complexity acoustic scene classification for multi-device audio:\\ analysis of DCASE 2021 Challenge systems }
\name{Irene Mart\'{\i}n-Morat\'o, Toni Heittola, Annamaria Mesaros, Tuomas Virtanen \thanks{This work was supported in part by the European Research Council under the European Unions H2020 Framework Programme through ERC Grant Agreement 637422 EVERYSOUND.}}
 \address{Computing Sciences \\ Tampere University, Finland\\
 \{irene.martinmorato, toni.heittola, annamaria.mesaros, tuomas.virtanen\}@tuni.fi}
\begin{document}

\ninept
\maketitle

\begin{sloppy}

\begin{abstract}
This paper presents the details of Task~1A Low-Complexity Acoustic Scene Classification with Multiple Devices in the DCASE 2021 Challenge. The task targeted development of low-complexity solutions with good generalization properties. The provided baseline system is based on a CNN architecture and post-training quantization of parameters. The system is trained using all the available training data, without any specific technique for handling device mismatch, and obtains an overall accuracy of 47.7\%, with a log loss of 1.473. The task received 99 submissions from 30 teams, and most of the submitted systems outperformed the baseline. The most used techniques among the submissions were residual networks and weight quantization, with the top systems reaching over 70\% accuracy, and log loss under 0.8. 
 
The acoustic scene classification task remained a popular task in the challenge, despite the increasing difficulty of the setup.
\end{abstract}


\section{Introduction}
\label{sec:intro}

Acoustic scene classification aims to classify a short audio recording into a set of predefined classes, based on labels that indicate where the audio was recorded \cite{Benetos2018}. The popularity of the task in DCASE Challenge throughout the years has allowed the development of approaches diverging from the textbook supervised classification, introducing along the years different devices \cite{Mesaros2018_DCASE}, open-set classification, and low-complexity conditions \cite{Heittola_2020_DCASE}, along with the publication of suitable datasets. 

Classification performance is easily affected by differences in recording devices, therefore solutions robust to device mismatch are important for real applications. Mismatches between training and testing data, typically present in real scenarios, have a significant effect on the generalization properties of deep learning models. A variety of solutions were proposed for dealing with the mismatch: for example in DCASE 2019 challenge, Kosmider et al. \cite{Kosmider2019} used a spectrum correction method to account for different frequency responses of the devices in the dataset, based on the fact that the provided development data contained temporally aligned recordings from different devices. Other systems used multiple forms of regularization that involves aggressively large value for weight decay, along with mixup and temporal crop augmentation \cite{Gao2019}.
In DCASE 2020 Challenge, most of the submitted systems used multiple forms of data augmentation including resizing and cropping, spectrum correction, pitch shifting, and SpecAugment, which seems to compensate for the device mismatch \cite{Heittola_2020_DCASE}. The top system had an accuracy of 76.5\% (1.21 log loss), using residual networks for the 10 scene classification with mismatched data \cite{Sangwon2020}.

In addition to dealing with data collected from devices not available during training, real-world solutions also need to be able to operate on devices with limited computational capacity \cite{Sigtia2016}.
For instance, in SED, dilated CNN have been applied to reduce the number of model's parameters \cite{Li2020} whereas in \cite{Gordon2018}, network dimensions have been scaled to obtain smaller and efficient architectures. In DCASE 2020, the low-complexity classification task consisted of a 3-class problem, to which many of the submissions imposed restrictions on the model architectures and their representations, such as using slim models, depth-wise separable CNNs, pruning and post-training quantization of model weights \cite{Heittola_2020_DCASE}. 
The top system \cite{Koutini2020} used a combination of pruning and quantization, using 16-bit float representation for the model weights and having a reported sparsity of 0.28 (ratio of zero-valued parameters), obtaining 96.5\% accuracy (0.10 log loss).

\begin{figure}
    \centering
    \includegraphics[width=0.9\columnwidth]{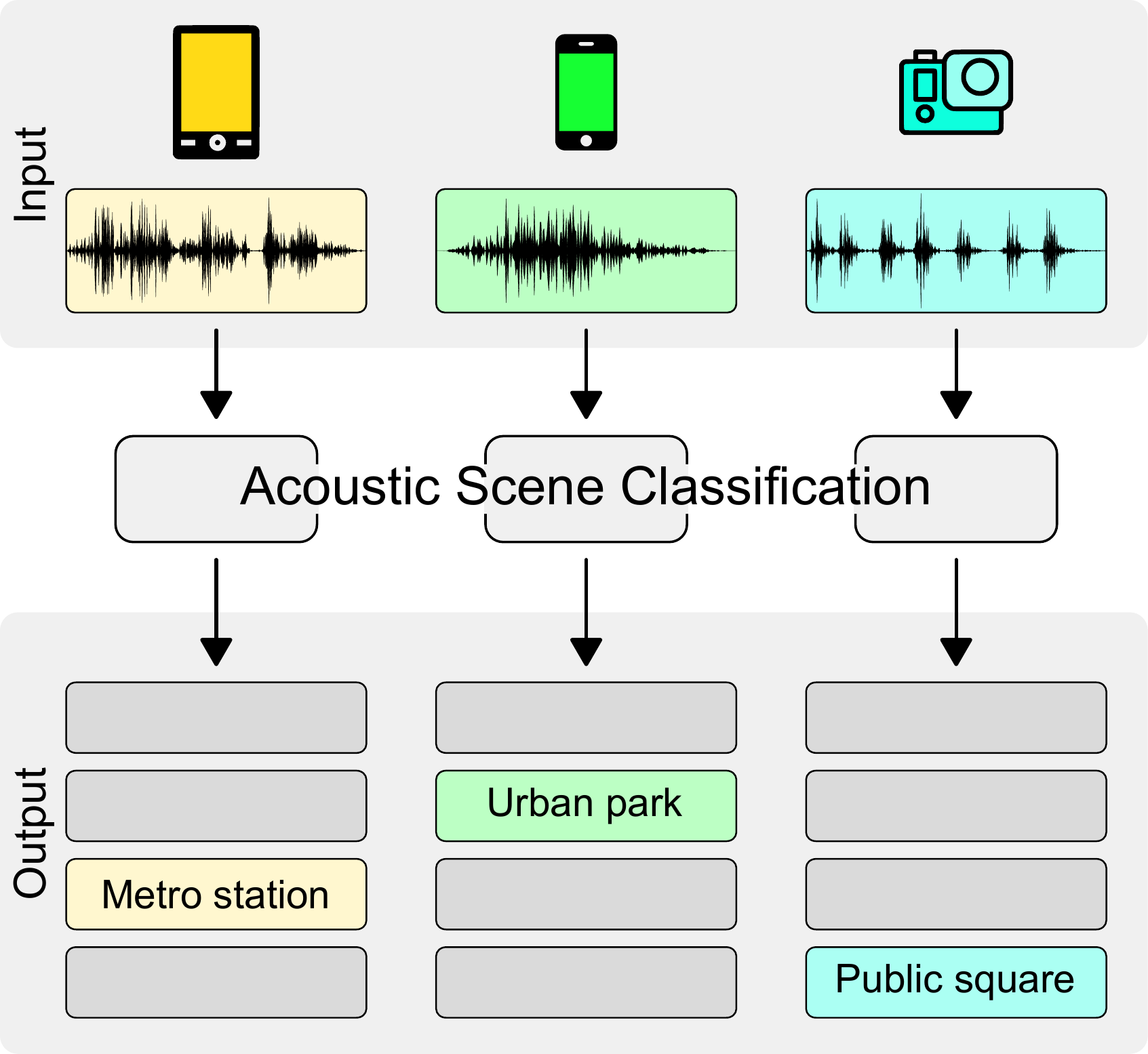}
        \caption{Acoustic scene classification for audio recordings.}
    \label{fig:overview}
    \vspace{-10pt}
\end{figure}

In DCASE 2021 Challenge, the two problems are merged, targeting good performance for a 10-class setup, multiple devices, and with model size constraints.
The added difficulty of the task comes from imposing more demanding conditions on both studied directions: using 10 classes instead of the three classes like in 2020, and dropping the model size limit from 500KB to 128KB. The choice of these conditions is motivated by the good performance demonstrated in DCASE 2020 Challenge, which showed that it was possible to achieve high performance results with a low complexity model.

This paper introduces the results and analysis of the DCASE 2021 Challenge Subtask A: Low-Complexity Acoustic Scene Classification with Multiple Devices. The paper is organized as follows: Section 2 introduces shortly the setup, dataset, and baseline system. Section 3 presents the challenge participation statistics in terms of numbers and use of methods, and Section 4 presents a detailed analysis of the submitted systems. Finally, Section 5 presents conclusions and thoughts for future development of this task.

\section{Task setup}
\label{sec:task-setup}

The specific feature of this task for acoustic scene classification is generalization across a number of different devices, while enforcing a limit on the model size. The 11 different devices in the dataset include real and simulated devices, and the model limit is 128 KB.

\subsection{Dataset and performance evaluation}

The task used the \textbf{TAU Urban Acoustic Scenes 2020 Mobile} dataset \cite{Dataset2020_Mobile_dev,Dataset2020_Mobile_eval}. The dataset is the same as used in DCASE 2020 Challenge Task 1A, comprised of recordings from multiple European cities, ten acoustic scenes \cite{mesaros_2019_DCASE}: \textit{airport}, \textit{indoor shopping mall}, \textit{metro station}, \textit{pedestrian street}, \textit{public square}, \textit{street with medium level of traffic}, \textit{travelling by a tram}, \textit{travelling by a bus}, \textit{travelling by an underground metro} and \textit{urban park}. Four devices used to record audio simultaneously are denoted as A, B, C, and D (real devices), with an additional 11 devices S1-S11 simulated using the audio from device A. The development/evaluation split, and the training/test split of the development set are the same as in the previous challenge, with 64 hours of audio available in the development set and 22 hours in the evaluation set.
For details on the dataset creation, and the amount of data available from each device, we refer the reader to \cite{Heittola_2020_DCASE}.

We evaluate the submitted system using multi-class cross-entropy and accuracy, calculated as macro-average (average of the class-wise performance for each metric). We use multi-class cross-entropy (log loss) for ranking the systems, in order to create a ranking independent of the operating point. Accuracy values are provided for comparison with the methods evaluated in previous years.

\subsection{System complexity requirements}

A model complexity limit of 128 KB was set for the non-zero parameters. This limit allows 32768 in float32 (32-bit float) representation, which is often the default data type (32768 parameter values * 32 bits per parameter / 8 bits per byte= 131072 bytes = 128 KB). Different implementation may consider minimizing the number of non-zero parameters of the network in order to comply with this size limit, or representation of the model parameters with a low number of bits. 

The computational complexity of the feature extraction stage is not included in this limit, due to a lack of established methodology for estimating and comparing complexity of different low-level feature extraction implementations. By excluding the feature extraction stage, we keep the complexity estimation straightforward across approaches. Some implementations may use a feature extraction layer as the first layer in the neural network - in this case the limit is applied only to the following layers, in order to exclude the feature calculation as if it were a separate processing block. However, in the special case of using learned features (so-called embeddings, like VGGish~\cite{Hershey2017}, OpenL3~\cite{Cramer_2019_ICASSP} or EdgeL3~\cite{Kumari_2019_IPDPSW}), the network used to generate them counts in the calculated model size.

\section{Baseline system}

The baseline system is based on a convolutional neural network (CNN) with the addition of model parameters quantization to 16 bits (float16) after training.  

The system uses 40 log mel-band energies, calculated with an analysis frame of 40 ms and 50\% hop size, to create an input shape of $40\times500$ for each 10 second audio file. The neural network consists of three CNN layers and one fully connected layer, followed by the softmax output layer. Learning is performed for 200 epochs with a batch size of 16, using Adam optimizer and a learning rate of 0.001. Model selection and performance calculation are done similar to the baseline system in DCASE 2020 Challenge Subtask A.
Quantization of the model is done using Keras backend in TensorFlow 2.0 \cite{abadi2016}, after training the model, the weights are set to \textit{float16} type.
The final model size of the system after quantifying is 90.3 KB.

\begin{table}[]
\centering
\begin{tabular}{r|c|c|c}
\toprule
System &  Log loss & Accuracy & Size \\
\midrule
keras & 1.473 ($\pm$ 0.051) & 47.7\% ($\pm$ 0.9) & 90.3KB \\
\bottomrule
\end{tabular}
\caption{Baseline system size and performance on the development dataset}
\label{tab:models}
\end{table}

\begin{table}[!b]
\centering
\begin{tabular}{r|c|c}
Scene label	&	Log Loss &	Accuracy\\
\midrule
Airport	          & 1.43  &  40.5\% \\
Bus	              & 1.32 & 47.1\% \\
Metro             & 1.32 & 51.9\% \\
Metro station     & 1.99 & 28.3\% \\
Park              & 1.17 & 69.0\% \\
Public square     &	2.14 &	25.3\% \\
Shopping mall     &	1.09 &	61.3\% \\
Pedestrian street &	1.83 &	38.7\% \\
Traffic street    & 1.34 &	62.0\% \\
Tram              & 1.10 &	53.0\% \\
\midrule
Avg.              & \textbf{1.473} & \textbf{47.7\%} \\

\end{tabular}
\caption{Class-wise performance of the baseline system on the development dataset.}
\label{tab:dev_result}
\end{table}

The classification results on the development dataset training/test split is presented in Table~\ref{tab:dev_result}. 
Given the results shown in this table, \textit{shopping mall} is the class with the lowest log loss, while public square has the highest log loss, being the most difficult to classify. The system behaves similarly to previous year challenge task 1A, the loss only increases 0.108 while the accuracy drops 6.4 points.

\begin{table*}[]
    \centering
    \begin{tabular}{c|l|cc|ccc|ll}
    \toprule
         Rank & System &  \makecell{Logloss\\$\pm95\%$ CI} & \makecell{Acc\\$\pm95\%$ CI$(\%)$} & \makecell{Size\\ (KB)} & Weights & Sparsity & Learning & Architecture \\
         \midrule
         1 & Kim\_QTI\_2 &  0.72$\pm$0.03 &  76.1$\pm$0.94 & 121.9 & int8 & \checkmark & KD & BC-ResNet  \\
         3 & Yang\_GT\_3 &  0.74$\pm$0.02 &  73.4$\pm$0.97 & 125.0 & int8 & \checkmark & KD & Ensemble \\
         9 & Koutini\_CPJKU\_3 & 0.83$\pm$0.03 & 72.1$\pm$0.99 & 126.2 & float16 & \checkmark & grouping CNN & CP\_ResNet  \\
         12 & Heo\_Clova\_4 &  0.87$\pm$0.02 & 70.1$\pm$1.01 & 124.1 & float16 & - & KD & ResNet \\         
         13 & Liu\_UESTC\_3 &  0.88$\pm$0.02  & 69.6$\pm$1.01 & 42.5 & 1-bit & -  & - & ResNet \\      
         17 & Byttebier\_IDLab\_4  & 0.91$\pm$0.02 & 68.8$\pm$1.02 & 121.9 & int8 & \checkmark & grouping CNN & ResNet \\    
         19 & Verbitskiy\_DS\_4  & 0.92$\pm$0.02  & 68.1$\pm$1.03 & 121.8 & float16 & - & - & EfficientNet  \\  
         22 & Puy\_VAI\_3 &  0.94$\pm$0.02   & 66.2$\pm$1.04 & 122.0 & float16 & - & focal loss & Separable CNN  \\  
         25 & Jeong\_ETRI\_2  & 0.95$\pm$0.03  & 67.0$\pm$1.04  & 113.9 & float16 & - & - & Trident ResNet  \\  
         28 & Kim\_KNU\_2  & 1.01$\pm$0.03  & 63.8$\pm$1.06 & 125.1 & int8 & - & mean-teacher & Shallow inception  \\ 
         \midrule
         85 & Baseline & 1.73$\pm$0.05 & 45.6$\pm$1.10 & 90.3 & float16 & - & - & CNN\\
         \bottomrule
    \end{tabular}
    \caption{Performance on the evaluation set and complexity management techniques for selected top systems (best system of each team). ``KD'' refers to Knowledge Distillation and ``BC'' stands for Broadcasting. }
    \label{tab:top10}
\end{table*}

\section{Challenge results}
\label{sec:results}

This section presents the challenge results and an analysis of the submitted systems. A total of 99 systems were submitted for this task from 30 teams. The number of participants for the ASC task is steady in the recent years, showing that its popularity does not decrease, but continues to attract attention through different setups. 

The highest accuracy obtained for the classification was 76.1\%, for the system of Kim\_QTI \cite{Kim2021b}, with the same system also having the best log loss of 0.724. Overall, 18 submitted systems had over 70\% accuracy.

The performance and a few selected  characteristics for systems submitted by the top 10 teams (best system of each team) are presented in Table \ref{tab:top10}. Confidence intervals for log loss were calculated using the jackknife estimation as in \cite{Mesaros2019}.

The ranking of systems is based on log loss; if the systems would be ranked by accuracy, their order would be quite different: while top 3 teams would keep their spots, teams ranked 12th and 27th would move to ranks 4th and 8th. In the systems ranking, systems ranked 1st, 2nd, 9th and 10th would keep their place, while the others in between would be shuffled. We calculated the Spearman rank correlation between accuracy and log loss, to investigate the strength of the association between the two. The correlation is 0.73, which, while strong, indicates quite significant changes in the ranking over the entire list of 99 systems.

\subsection{Features and augmentation techniques}

All top 10 teams make use of mel energies as feature representation, ranging from 40 to 256 mel bands, with three of them adding also delta and delta-delta values. Overall, only three out of 30 teams do not use log mel as input features; instead, they use gammatone (Naranjo-Alcazar\_ITI), deep scattering spectrum (Kek\_NU) and embedding from AemNet (Galindo-Meza\_ITESO). Augmentation techniques are also used, with most popular techniques being mixup (used by 20 teams) and specAugment (10 teams). Other augmentation techniques used, known as label-invariant transformations, are pitch shifting, temporal cropping or speed change, and they are commonly used to improve the performance of CNN networks.

\subsection{Architectures}

Residual models are the most popular ones; in fact a total of 15 teams use them, among them the top five models, with the exception of the second best team, Yang\_GT, which uses ensembles of CNNs. In the literature there are only a handful of models suitable for usage with devices constrained by processing power and/or memory. Some of these models are MobileNet \cite{Sandler2018} and EfficientNet \cite{Tan2019}, which are networks based on residual blocks. The most recent one, EfficientNet, also contains squeeze-and-excitation blocks. A total of five teams used some modified version of such models. Finally, the two models that perform below the baseline accuracy make use of full 2D-CNN models.

\begin{figure*}
    \centering
    \includegraphics[width=0.9\textwidth]{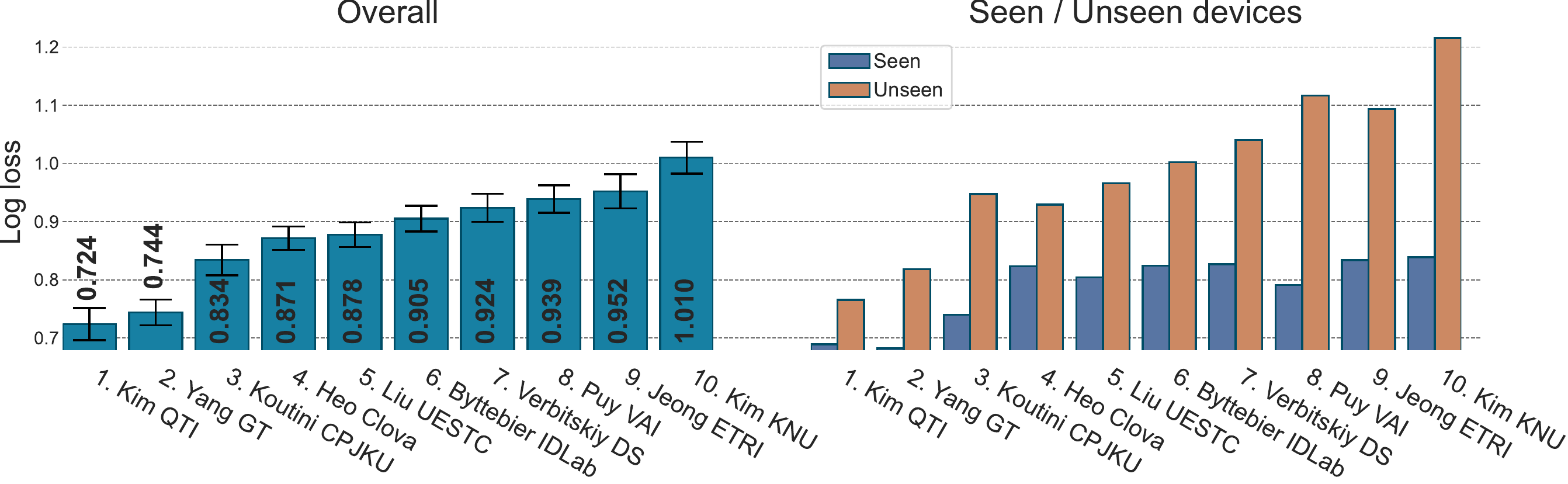}
    \caption{Classification log loss for the 10 top teams (best system per team) on the evaluation dataset.}
    \label{fig:logloss}
\end{figure*}

\subsection{System complexity}

Regarding model complexity, the top 10 systems, belonging to three different teams, Kim\_QTI \cite{Kim2021b}, Yang\_GT \cite{Yang2021} and Koutini\_CPJKU \cite{Koutini2021}, are close to the allowed model size limit. They range from 110~KB to 126.81~KB, with the system ranked first having a size of 121.9KB. We have to go down to position 77 (1.464 log loss, 47.17\% accuracy) to find the smallest model of 29~KB by Singh\_IITMandi, which used a filter pruning strategy consisting of 3 steps and one extra for final quantization of the weights to 16-bits. 

A notable small model, with size 42.5KB, belongs to a top 5 ranked team, Liu\_UESTC \cite{Liu2021}. This specific system is ranked 13th, with a 0.878 log loss and 69.60\% accuracy. The model compression is performed with 1-bit quantization, similar to the McDonnell\_USA system from DCASE2020 Challenge Task 1B \cite{Gao2019}. Despite the high performance in DCASE2020, this is the only team using the one-bit quantization approach this year. 

There are only two teams that do not use any quantization: Pham\_AIT \cite{Pham2021} uses channel restriction and decomposed convolution, while Qiao\_NCUT does not mention any quantization; however, these are not in the top 10 ranked teams. On the other hand, 11 teams perform pruning with some quantization technique, and two teams perform the Lottery Ticket Hypothesis (LTH) \cite{Frankle2019} pruning method. One achieved second position, with a model of size 125KB, while the other stayed below the baseline with a model size of 124KB. The main difference between the two is the use of ensemble of CNN with knowledge distillation vs a single CNN model. 

Therefore, sparsity used in combination with quantification is a very popular and efficient way of reducing the model size; however, model architecture and other learning techniques have to be taken into account in order to achieve good classification performance.

\subsection{Device and class-wise performance}

All systems have higher performance on the devices seen during training (A, B, C, S1, S2, S3) than on the unseen ones (D, S7, S8, S9, S10), with a difference in accuracy of almost 3\% (statistically significant) for the system ranked first. As seen in Figure \ref{fig:logloss}, this difference increases as we go down the system ranking, reaching an almost 10\% gap when considering accuracy, and 0.37 when considering log loss, for team Kim\_KNU. The Spearman's rank correlation between the accuracy on seen and accuracy on unseen devices is 0.92, while between the log loss on the seen devices and the log loss on the unseen devices is 0.91. These values indicate that while they are very highly correlated, the gap between the two does not always preserve the ranking order. 

The generalization properties of the systems are worst regarding the unseen devices, while for seen/unseen cities the performance does not vary as much. Some systems get better performance for unseen cities. Indeed, the correlation between performance on seen cities and on unseen cities is 0.95, while device-wise is 0.91.  
This indicates that data mismatch due to the unseen devices is more challenging than the mismatch created by different cities, due to the different properties of the recorded audio, which are related to the device-specific processing. In particular, the poor performance on the unseen devices is mostly due to device \textbf{D}, which is the GoPro, while the other devices are real and simulated mobile phones and tablets, developed with closer attention to the voice/audio transmission quality. Indeed, we can see that accuracy on device \textbf{D} is the lowest one on average (48.66\%) while device \textbf{A} reaches an accuracy of 72.45\%.

The most difficult to classify are \textit{airport} and \textit{street pedestrian} classes, while the easiest to classify is \textit{street traffic} with 80\% acc. and 0.283 log loss on average for all the systems. Among the techniques used for increasing the generalization capabilities we can find
residual normalization \cite{Kim2021b}, domain adversarial training \cite{Koutini2021}, and use of data augmentation techniques as performed in \cite{Yang2021, Hee-Soo2021}.

\subsection{Discussion}

Residual Networks have been shown to be the most efficient regarding acoustic scene classification for complexity-constrained solutions. Quantization combined with sparsity techniques have kept the model complexity within the required limit. The solutions presented in DCASE2021 Challenge follow the trends from previous year, combining the best characteristics and techniques from both acoustic scene classification subtasks. It is proven that the use of data augmentation improves generalization, compensating device mismatch. However, the reported log loss for seen/unseen devices and cities, shows that there is room for improvement; e.g. the use of domain adaptation techniques, like adversarial training used in \cite{Koutini2021}, is not sufficient to deal with mismatches, since they report the highest mismatch among the 10-best submissions, while the use of mixup techniques prove to be more efficient. 

Other mechanisms with less direct impact on the model parameters can be applied during the training step, the so-called learning techniques. These algorithms focus on obtaining a more efficient model by training it differently. Among the submissions, half of the teams have made use of some version of these techniques, the most popular ones being the use of focal loss and knowledge distillation. Focal loss helps the model during training to pay attention to the more difficult samples. However, the use of knowledge distillation seems to be the more efficient one, considering the ranking of related solutions.

\section{Conclusions and future work} 
\label{sec:conclusions}

This paper presented the results of the DCASE 2021 Challenge Task~1A, Low-Complexity Acoustic Scene Classification with Multiple Devices. The task combines the need for robustness and generalization to multiple devices of such systems with the requirement for a low-complexity solution, bringing the research problem closer to real-world applications. The method for calculating the model complexity includes only the parameters of the network, with exceptions in the case of employing embeddings. However, the strict model size limit has rendered use of embeddings impossible, as most currently available pretrained models are already too big for the imposed limit.
The task has received a large number of submissions that brought into spotlight interesting techniques that combine the best performing methods from the point of view of robustness, like data augmentation, with methods directed towards obtaining light models, e.g., knowledge distillation, weights quantization, and sparsity. The popularity of the task shows that acoustic scene classification is still relevant for the audio community, and in particular, the development of solutions applicable for real-life devices.

% -------------------------------------------------------------------------
% Either list references using the bibliography style file IEEEtran.bst

\bibliographystyle{IEEEtran}
\bibliography{refs}

\end{sloppy}
\end{document}